\documentstyle[12pt,psfig,epsf]{article}
\newcommand{\be}{ \begin{eqnarray}}
\newcommand{\ee}{\end{eqnarray}}
\newcommand{\beno}{ \begin{eqnarray*}}
\newcommand{ \eeno}{\end{eqnarray*}}
\newcommand{\raf}[1]{(\ref{#1})}

\begin{document}

\bibliographystyle{/homeb/vkoch/tex/bst/try}
 \begin{titlepage}
\hspace{11cm}
{\large LBL--38619}

\hspace{11cm} 
{\large UC--413}

%\hspace{11cm}
%{\large nucl-th/9512029}
\vspace{.7cm}
\begin{center}
\ \\
{\large {\bf Dilepton Production at SPS-Energy Heavy Ion
Collisions}}\footnote{Dedicated to Gerry Brown in honor of the $32^{nd}$
celebration of his $39^{th}$ birthday.}\footnote{This work was supported
 by the Director, 
Office of Energy Research, Office of High Energy and Nuclear Physics, 
Division of Nuclear Physics, Division of Nuclear Sciences, of the U.S. 
Department of Energy under Contract No. DE-AC03-76SF00098.}
\vspace{2cm}
\ \\
{\large Volker Koch and Chungsik Song}
\ \\
\ \\
{\it Nuclear Science Division, Lawrence Berkeley National Laboratory,\\
University of California,\\
Berkeley, CA, 94720, U.S.A.}\\
\ \\
%\today \\
\ \\

\vspace{2cm}
{\large \bf Abstract}\\
\vspace{0.2cm}
\end{center}
\begin{quotation}
The production of dileptons is studied within a hadronic transport model. We
investigate the sensitivity of the dilepton spectra to the initial 
configuration of the hadronic phase in a ultrarelativistic heavy ion collision.
Possible in medium correction due to the modifications of pions and the pion
form factor in a hadronic gas are discussed.
\end{quotation}
\end{titlepage}
\section{Introduction}
\label{sec:0}
One of the major goals of the ultrarelativstic heavy ion program is 
to study the
restoration of chiral symmetry at high temperatures and densities and possibly
to create and identify a new form of matter the so called Quark Gluon Plasma. To
this end the measurement of electromagnetic probes such as photons and 
dileptons have continually attracted great interest. Contrary to hadronic
observables these weakly 
interacting probes are not affected by final state interactions and thus may 
provide information about the early, hot stage of the reaction. 
While at very high invariant masses, above that of the $J/\Psi$ the dilepton
yield is essentially dominated by the Drell-Yan production from the hard 
nucleon nucleon collisions, it has been suggested  that in the mass region
between the $\Phi$ and the $J/\Psi$, dileptons originating from the QGP can be
observed (for an overview see e.g. ref. \cite{Rus92}).    
Of particular interest in the high invariant mass region has been 
the study of the suppression of the $J/\Psi$, because of its possible 
implications about deconfinement. 
At low invariant masses, below $1 \, \rm GeV$ the dilepton spectrum is
expected to be dominated by 
the hadronic phase, where processes such as hadronic 
decays and pion annihilation contribute. 
This region is of particular interest, because possible 
in medium modifications of hadron masses such as that of vector mesons
may be observable in the dilepton invariant mass spectrum 
\cite{K92,WCW93,LK95}.
This mass region is also accessible at lower bombarding energies (BEVALAC, SIS)
where modifications of hadron masses due to finite nuclear density rather than
temperature can be studied \cite{DLS95,WCW93}

Recently interest \cite{LKB95,CEK95,RCW95,SKL96,SSG96,Hag96} 
in the low mass region has been sparked by the experimental
data from the CERES collaboration \cite{CERES95}, which show a considerable
enhancement at invariant masses of about $400 \, \rm MeV$ for SPS-energy 
S+Au heavy ion
collisions as compared to proton-proton as well as proton-nucleus collisions.
A similar enhancement has also been seen by the HELIOS collaboration at more
forward rapidities \cite{HELIOS95}. 
It has been suggested \cite{LKB95,CEK95} 
that this enhancement may be due to the dropping of the vector meson masses 
at finite temperature and densities, as proposed  in 
\cite{Pis82,BR91}. However, also more conservative effects, such as
a modified in medium pion dispersion relation may give sufficient enhancement
\cite{SKL96,RCW95}. This later effect has first been suggested in the context
of BEVALAC energy heavy ion collisions \cite{GK87,KP90,KXK90}
where the pion dispersion relation is modified due to the interactions with
nucleons forming deltas. At SPS energies, one would expect
that the major modification should come from interactions among pions, mostly 
in the the isovector - p-wave channel, 
which is dominated by the s-channel rho-meson resonance
\cite{Shu91,Son94}. The effect of the nucleons, however, may not be negligible 
\cite{RCW95}.

However, before definite conclusions about possible in medium effects of any
kind can be drawn, a conventional explanation of the observed enhancement has
to be ruled out. The cocktail used by the CERES collaboration to estimate the
conventional background includes only the decay of hadrons with a relative
abundance taken from proton proton data and scaled with the number of charged
particles.  In a heavy ion collision several things could be different from
these assumptions due to multiple elastic and inelastic scatterings. First of
all the relative abundance could be altered due to creation an destruction of
higher lying meson resonances such  the $\rho$ or the $a_1$. Secondly, there
could be multiple $\pi$-$\pi$ annihilation processes leading to additional
dilepton yields. Furthermore, initially , after formation, the hadrons not
necessarily have to be in thermal equilibrium.  One rather could imagine that a
great deal of the thermalization observed in the final (freeze out) state
\cite{BSW96} actually takes place within the hadronic phase. 
This equilibration, 
which involves elastic and inelastic collisions, then may result in a quite
different dilepton yield as, for instance, obtained from an equilibrated hadron
gas. In other words, a careful and systematic investigation of the dilepton
production may reveal information about the equilibration properties of the
hadronic phase. 

It appears rather unlikely, on the other hand,
that the equilibration process can be inferred from 
hadronic observables only. Provided that a certain degree of equilibration has
been obtained in the final state, as suggested by a recent analysis of AGS as
well as SPS data \cite{BSW95,BSW96}, it is of course impossible to 
extract in which way the system has reached this point. Penetrating probes,
however,  are not only sensitive to the final state but represent an
integral over the entire reaction history and thus are much better suited to
constrain our understanding of equilibration dynamics. 
In other words, the fact the
event generators such as RQMD \cite{RQMD} or VENUS \cite{VENUS} are able to
reproduce most hadronic observables not necessarily implies that they simulate
the reaction dynamics properly, since there are infinitely many trajectories
towards thermal equilibrium.  

The purpose of this article is to investigate the sensitivity of the dilepton
yield to different initial conditions in the hadronic phase and to see to which
extent the CERES and HELIOS data can be accounted for by non equilibrium
processes in the hadronic phase. Our strategy therefore is to parameterize the
initial phase space and the abundances of the hadrons subject to the
constraint, 
that the final hadronic spectra and rapidity distributions are reproduced.
This will lead to a set of initial conditions for
which we then study the contribution to the dilepton data.

This paper is organized as follows: In the first section we will describe the
hadronic transport model. In the second section we will describe the initialization of the
hadronic phase space. In the third section  the production of
dileptons within out transport model will explained. 
We then will turn to the discussion of
possible medium effect before we present the results and our prediction for
Pb+Pb collisions.

\section{Transport Model}
\label{sec:1}
In this section we will describe the hadronic transport model used to study
the dilepton production. In our calculation we include pions, $\rho$- $\omega$-
and $a_1$ mesons as well as nucleons and deltas for the baryons. We ignore
higher baryon resonances at this time but we will present an estimate of their
contribution to the dilepton spectrum below. Since the fraction of baryons in
the relevant rapidity region is small compared to that of 
the pions, we do not expect
that they contribute significantly to the equilibration process. We have also
left out the strange mesons for the same reason. The contribution of the eta
meson to
the dilepton yield is calculated at the end using the measured eta/pi ratio
\cite{WA80_94}. 

The interaction between nucleons, deltas and pions is done in the standard
fashion as reported e.g. in \cite{BKM93,DB91,BD88} and we refer the 
reader to these references for details. 

All hadronic resonances are propagated explicitly. For instance the
resonant isovector p-wave  $\pi - \pi$ scattering in the $\rho$ channel is
split into the formation, the propagation and the decay of a $\rho$-meson. As a
consequence the mass of these resonances are not restricted to their central
value but may  assume any value as determined by the production process
of by the initialization procedure (see below). To be specific, for the 
formation of the $\rho$ from $\pi - \pi$ collisions, we use an isospin averaged
Breit-Wigner cross section
\be
\sigma_{\pi \pi \rightarrow \rho}(M)  = \frac{8 \pi}{p^2} 
\frac{\Gamma^2_{\rho}(M) M^2}{ 
(m^2 - m_\rho^2)^2 + M^2 \Gamma^2_{\rho}(M) }
\ee
Here $p$ is the center of mass
momentum and $M$ the invariant mass of the pair. Using the standard
$\pi$-$\rho$ coupling \cite{Bha88}  the width of the $\rho$ is given by

\be
\Gamma_{\rho \rightarrow \pi \pi}(M) =  \Gamma_{0,\rho} \frac{p^3}{p_0^3} \frac{m_\rho^2}{M^2}
\ee
where $p$ is the decay momentum. $\Gamma_{0,\rho} = 150 \, \rm MeV$ and $p_0$
are the width and the decay momentum at peak value ($M= m_\rho$).

The formation and decay of the $a_1$ is a little more complicated since it
involves a hadronic resonance, the $\rho$, in the initial or final state,
respectively. Thus, the $a_1$ does not necessarily decay into a $\rho$ of mass
$m_\rho = 770 \, \rm MeV$ but also into states with smaller and larger
masses. As a starting point, we use the formula of reference \cite{XSB92} for
the decay width of the $a_1$ into pion and $\rho$
\be
\Gamma_{a_1}(M,m_\rho) =  
\frac{G_{a_1 \rho \pi}^2 p}{24 \pi
m_{a_1}^2} \left( \frac{1}{2} (M^2 - m_\rho^2 - m_\pi^2)^2 + m_\rho^2 (m_\pi^2
+ p^2) \right)
\label{a-width1} 
\ee
Notice, that in the above formula the width depends also on the mass of the
outgoing $\rho$. Therefore, if we want to allow for the decay into
$\rho$-mesons with masses different from the central value the total
$a_1$-width should be given by the integral of the above formula over all
kinematically allowed $\rho$-masses weighted with the $\rho$ mass distribution
$f_\rho(m)$.
Hence
\be
\Gamma_{a_1}^{tot}(M) = \int_{2 m_\pi}^{\infty} f_\rho(m_\rho) 
\Gamma_{a_1}(M,m_\rho)
\ee
with
\be
f_\rho(m) = \frac{2}{\pi} \frac{m^2 \Gamma_\rho(m)}{ (m^2 - m_\rho^2)^2 + m^2
\Gamma_\rho^2}
\label{a-width1_tot} 
\ee
We thus have to adjust the coupling $G$ in eq. \raf{a-width1} such that the
total width $\Gamma_{a_1}^{tot}$ agrees with the experimental value of $400 \,
\rm MeV$. We find $G_{a_1 \rho \pi} = 18.1 \, \rm GeV^{-2}$. 
The {\em partial} width for the decay into a $\rho$ of a specific
mass is then given by
\be
\frac{d \Gamma}{d m_\rho} = f_\rho(m_\rho) \Gamma_{a_1}(M,m_\rho)
\ee
This then gives the distribution from which the mass of the outgoing $\rho$ has
to be picked.
Finally, the isospin averaged cross section for the formation of an $a_1$ in a
$\pi - \rho$ collision is given by
\be
\sigma_{\pi \rho \rightarrow a_1}(M) = \frac{3 \pi}{4 p^2} 
\frac{M^2 \, \Gamma_{a_1}(M,m_\rho) \Gamma_{a_1}^{tot}(M) }{ 
(M^2 - m_{a_1}^2)^2 + M^2 {\Gamma^{tot}_{a_1}}^2(M) }
\ee
where $\Gamma_{a_1}(M,m_\rho)$ is given by eq. \raf{a-width1} and  
$\Gamma_{a_1}^{tot}(M)$ is given by eq. \raf{a-width1_tot}.
The advantage of this somewhat complicated prescription is that it allows to
include the finite width of the $\rho$ and, at the same time, preserves
detailed balance. 

For the $\omega$ meson we adopt a similar strategy. We model the dominant decay
into three pions as a two step process via an intermediate $\pi \, \rho$ state,
as suggested by chiral models including vector mesons \cite{Son93}. Thus in
our approach the $\omega$ decays into a pion and a $\rho$-meson 
the mass of which 
is below the central value. As explained previously in our transport
model these $\rho$'s are treated as real particles. To preserve detailed
balance we also allow for the formation of the $\omega$ by $\pi \,
\rho$-fusion. 

Following ref \cite{Son93} we use a p-wave form for the decay width of the
$\omega$ 
\be
\Gamma_\omega(m_\rho) = C q^3
\ee
where $q$ is the decay-momentum of the omega in its rest frame. The constant of
proportionality, $C$, 
is then determined such that the total width, integrated of all
possible $\rho$ - masses with the proper weighting, corresponds to the observed
experimental value for the decay into three pions
\be
\Gamma_\omega^{tot} = C \int_{2 m_\pi}^{m_\omega - m_\pi} \, 
f_\rho(m_\rho) \, q^3 \, dm_\rho = 7.4 \, \rm MeV
\ee
which gives a value of $C = 48 \, \rm  GeV^{-2}$.

Following the same arguments as in case of the $a_1$ the isospin averaged 
cross section for the formation of the $\omega$ is then given by
\be
\sigma(M,m_\rho) = \frac{4 \pi}{9 q^2} 
\frac{\Gamma_\omega(m_\rho)\, \Gamma_\omega^{tot}}{(M^2 - m_\omega^2) + M^2 
{\Gamma_\omega^{tot}}^2}
\ee

Naturally, due to the small width of the $\omega$ very few new $\omega$ are
being formed in the course of the expanding hadronic system. Therefore, the
resulting dilepton spectrum is only slightly changed as compared to
calculations which do not take the formation of the omega into account
\cite{LKB95}. 

Finally, for the non resonant scatterings we assume a constant cross section of
$20 \, \rm mb$ except for that among nucleons and Deltas, where a
parameterization of the measured nucleon-nucleon cross section is being used.

\section{Initialization of initial Phase space}
\label{sec:2}
As already outlined in the introduction, in our transport model 
we will not describe the entire
collision history, starting from the initial colliding nuclei. 
We rather want to
ask the question about the sensitivity of the dilepton production to the
hadronic phase-space 
configuration created in such a collision. Thus the initial stage in
our description is that of a hadronic gas with a possible non equilibrium
configuration in momentum space. We also do not impose chemical
equilibration. We, however, require that measured hadronic spectra and rapidity
distributions are reproduced at the end of the evolution. As we shall see this
leaves quite a number of possibilities to initialize the system.

\subsection{Configuration Space:}
We will assume that initially all particles are distributed uniformly  within
a cylinder of radius $R_0$ and longitudinal extent $2 Z_l$. This somewhat
simplified prescription may be improved by e.g. using a Gaussian distribution
in the longitudinal direction, which we do not consider here. However, we
should point out that the introduction of a formation time, discussed below,
will effectively introduce some smearing of the particle density in the
longitudinal direction. Since we plan to compare with the CERES data for
$S+Au$ collisions, we will assume an
initial transverse radius of $R_0 = 3.5 \, \rm fm$. 
The longitudinal extent will
be varied and the specific choices will be given in the results section 
\raf{sec:4}.

\subsection{Momentum Space:}
Momenta in the transverse direction are distributed according to a two
dimensional Bose- or Fermi- distribution with the possibility of non vanishing
chemical potential. 
The chemical potential of the $\rho$ is then twice that of
the pion and those of the $\omega$ and $a_1$ three times the pionic one.

For the longitudinal momenta we assume that the rapidities of the particles are
distributed according to a Gaussian distribution, 
\be
f(y,z) = \frac{1}{\sqrt{2 \pi \sigma^2}}  
\exp\left[ \frac{-(y - \frac{Y_l}{Z_l}z)}{2 \sigma^2} \right]
\label{eq:y_z}
\ee
which allows to take into account possible correlations between rapidity and 
longitudinal position $z$. To some extent, correlations between the
longitudinal position and the rapidity are already generated by the
introduction of a formation time as we will discuss below.

For the hadronic resonances the initial distribution of the masses has to be
specified as well. In principle there are two possibilities. First, the masses
are distributed according to the respective Breit Wigner distribution
$f_{hadron}(m)$. Second, assuming that thermal equilibration has been
achieved prior to hadronization, the masses should be distributed according to 
thermally weighted distribution, which slightly favors lower mass states.
Assuming boost invariance on gets
\be
f_{thermal}(m)= \frac{1}{N} f_{hadron}(m) \, m^2 T K_2(m/T)
\label{eq:mdis}
\ee
where $T$ is the transverse temperature, $K_2$  is the modified 
Bessel-function, and $N$ is a normalization factor. Notice, that one obtains
the same mass distribution in case of a Bjorken type fire cylinder. 
In this work we will only
consider the second possibility \raf{eq:mdis}.    
 
\subsection{Formation Time:}
A realistic description of the initial hadronic fireball should include a
formation time. Particles with higher rapidities will be created at later times
in a given frame of reference. In our approach we assume that particles will 
propagate {\em  without} interaction for times smaller than their
formation time
\be
t_{formation} = \tau \cosh(y)
\ee
where y is the rapidity of the particle under consideration and $\tau$ denotes
the proper formation time. This formation time clearly introduces a smearing of
the longitudinal distribution in configuration space in the sense that, at the
time particles will be allowed to interact, the fastest 
ones have already moved a
considerable distance $\delta z \simeq t_{formation} \, c$. For the same reason
the formation time leads to a rapidity  longitudinal-position correlation.

\noindent

\subsection{Particle Abundances:}
In order to reproduce the measured rapidity spectrum of protons by the NA35
collaboration \cite{NA35} we assume that our hadronic fireball contains 60
baryons. The relative weighting is then determined by the temperature assuming
thermal and chemical equilibrium. At the temperatures considered here we will
have about as many nucleon as deltas.

For the mesons we will either use the relative abundances as determined from
proton proton collisions, i.e $N_\rho / N_\pi = 0.2$, 
$N_\omega / N_\rho = 1/3$. Since the number of $a_1$ is not given by
experiment we assume $N_{a_1} / N_\omega = 0.5$.

Alternatively, we will assume initial chemical equilibrium and will
determine the relative abundances assuming a boost invariant momentum
distribution. In this case the abundances of course depend on the initial
temperature and will therefore be given in the appropriate context in the
results section \raf{sec:4}.

\subsection{Flow:}
Several analyses of available SPS data suggest that there is substantial
radial flow, which leads to larger apparent temperatures 
\cite{SSH93,BSW96}. We, therefore, allow for the possibility that the hadrons
have some transverse flow in the initial state. For simplicity we assume a
constant flow velocity independent of the radial distance of the particle. To
generate the flow, we first distribute the particles thermally in the
transverse direction, then we boost them according with the flow velocity
before we finally boost them in the longitudinal direction as given by the
initial rapidity distribution. This sequence of boost is identical to that
employed by Schnedermann et al. \cite{SSH93}. Allowing for transverse flow
reduces the initial temperature needed to reproduce the final transverse
momentum spectra. We should point out, however, that in our transport model
a certain amount of flow is generated as result of particle
collisions. Therefore, the initial flow velocities used here should not be
compared directly with results from the analysis of hadronic spectra.

\section{Dilepton production}
\label{sec:3}

Having specified the initial phase space distribution as well as transport
model we finally turn to the production of dileptons. 
As far as dilepton production through pion annihilation is concerned there are
essentially two ways to proceed in a transport model, which are in principal
equivalent. One is to directly calculate the dilepton yield from the pion
collisions and the other one is to obtain the dileptons from the decay of the
$\rho$ which has been formed in the collisions. Both are equivalent as has been
shown in \cite{LK95,LKB95}. 
However, we found that the second prescription is extremely
sensitive to statistical fluctuations for invariant masses smaller that $400 \,
\rm MeV$, as we will explain below. We, thus, resort to the first prescription,
namely the production of dilepton directly from the $\pi \, \pi$ collisions. 
We use the well known cross section
\be
\sigma_{\pi+\pi \rightarrow e^+ e^-}(M) = \frac29 \frac{4 \pi}{3} \frac{\alpha}{M^2}
\sqrt{1 - \frac{4 m_\pi}{M^2}} \frac{m_\rho^4}{(M^2 - m_\rho^4)^2 + M^2
\Gamma_\rho(M)^2}
\label{pi_pi-ann}
\ee
where the factor $\frac29$ arises from isospin averaging.

All the Dalitz decays are done explicitly. Therefore, in order to avoid
double counting, only the $\rho$ mesons present in the initial state are
allowed to decay into dileptons. All secondary $\rho$ mesons come from
processes such as pion annihilation or $a_1$ decay, where we take the
contribution to the dilepton yield into account explicitly, including 
vector dominance, so that those
$\rho$ mesons should not decay into dileptons. 

We furthermore include the direct decay of the $\omega$ into dileptons as well
as the Dalitz-decay of the $\eta$ meson. As already pointed out previously, we
do not explicitly propagate the $\eta$, but include it at the end of the
calculation. To that end we assume that the etas have the same rapidity
distribution as the pions. The relative abundance and the transverse momentum
spectrum we obtain from the measured $\eta/\pi_0$ ratios as a function of $p_t$
provided by the WA80 collaboration \cite{WA80_94}. 

In order to improve statistics we treat the dilepton decay `perturbatively',
i.e. at each time step the particles are allow to 'radiate' dileptons, without
being removed. The contribution to the final spectrum is then given by the
partial width multiplied by the time step size $\Gamma_{e^+e^-} \, dt$.
At the end of the calculation all remaining hadrons are forced to decay and
their contribution to the dilepton spectrum is given by the respective
branching ratios. 
In case of the pion annihilation, any $\pi \, \pi$ collision contributes to the
dilepton yield with the ratio of dilepton production cross section over total
cross section. 

For a comparison with the CERES data, cuts on the lepton momenta have to be
performed. In order to improve our statistics further we sample the decay of
each virtual photon several (ten) times.

For the Dalitz decays of the $\eta$ and  $\omega$  we use the formulae of
Landsberg \cite{Lan85} including form factors. For the form factors we use
the values given in \cite{Lan85}
\be
f_\eta (m) &=& (1 - 2.4 \, m^2)^2
\\
f_\omega (m) &=& (1. - m^2 /0.72^2)^2
\ee
where the invariant mass m is to be given in GeV.
 
For the $a_1$ we also use the Landsberg formula with a vector dominance form
factor including the width of the $\rho$
\be
f_\rho(m) = \frac{m_\rho^4}{(M^2 - m_\rho^4)^2 + M^2
\Gamma_\rho(M)^2}
\ee
We have compared this formula with an
explicit calculation of the Dalitz decay of the $a_1$ using
the Lagrangian of \cite{XSB92} an found very good agreement. 

For the partial decay width of the $\rho$ we use  \cite{Bha88}
\be
\Gamma_{\rho_0 -> e^+ e^-}(M) = \Gamma_0 \frac{m_\rho^3}{M^3}
\ee
where $\Gamma_0 = 6.77 \, \rm keV$ is the measured width at peak
value. 
Notice, that for small invariant masses the branching 
ratio for a $\rho$ decaying
into dileptons increases rapidly since the total width, which is dominated by
the p-wave decay into pion becomes small. 
As a consequence if an initial $\rho$ with small invariant mass 
happens to survive until the end of the
calculation it will decay with a very large branching ratio. We find that for 
400 runs the average number of $\rho$ with small invariant 
masses surviving until the end ($30 \, \rm fm/c$) is less than one and 
thus the statistical fluctuations are tremendous. By changing the random number
seed the results for invariant masses smaller that $400 \, \rm MeV$ fluctuated
by more than an order of magnitude. Since this mass region is identical to the
one where the CERES collaboration reports a strong enhancement, these
uncertainties are unacceptable. We should also point out that we have observed
similar fluctuations when we calculated the pion annihilation in a two step
process, i.e. via the formation and the decay of the $\rho$ meson. To avoid
those we resorted to the direct production of dilepton from pion collisions as
described above. 
In order to circumvent the statistical fluctuations from the  $\rho$ mesons in
the initial state we calculate the contribution to the dilepton spectrum from
$\rho$ mesons with masses smaller than $500 \, \rm MeV$ directly, i.e. not by
Monte Carlo methods. To this end we
calculate the number $N_\rho(M)$ of $\rho$ meson in a given mass interval 
$M + \Delta M$ from the mass and phase space distribution. The dilepton
yield is then obtained by multiplying with the branching ratio
\be
\frac{d N_{e^+e^-}}{d M} = N_\rho(M) 
\frac{\Gamma_{\rho \rightarrow e^+e^-}}{\Gamma_{tot}}
\ee
Since $N_\rho(M)$ is proportional to $\Gamma_{tot}$ the above distribution is
smooth even if $\Gamma_{tot}$ becomes small.
By calculating the decay of these small mass $\rho$ mesons directly we ignore
possible absorption effects and thus somewhat overestimate the yield at low
invariant masses. Naturally, in order not to double count, all initial $\rho$
mesons with masses below $500 \, \rm MeV$ are not allowed to decay in
dileptons during the expansion. 

\section{Medium modifications}
\label{sec:3b}

One of the  most interesting aspects of studying low mass dilepton pairs is the
possibility to learn something about in medium modifications of hadrons.
One possibility is to see the possible dropping of the mass of vector mesons
\cite{K92,LK95} as proposed by Pisarski \cite{Pis82} as well as Brown and
Rho \cite{BR91}. In the context of the CERES data, this possibility has been
studied in references \cite{LKB95} as well as \cite{CEK95}.
However, as far as the restoration of chiral symmetry is concerned, the
dropping of the vector meson masses are at best an indirect signature. To
leading order in the temperature chiral symmetry predicts no change in the mass
of the vector mesons \cite{DEI90}. This finding is also confirmed by
calculations using effective chiral Lagrangians
\cite{Son93b,Pis95}. 

\begin{figure}[tbh]
\setlength{\epsfysize}{3in}
\centerline{\epsffile{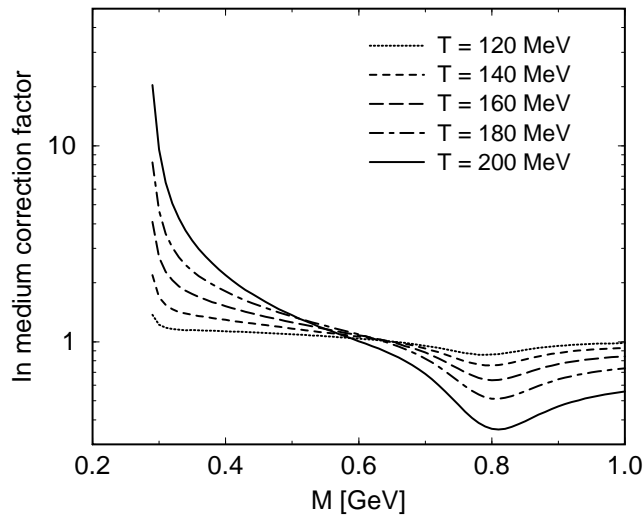}}
\caption{Correction factor used in in medium calculation for the pion
annihilation cross section}
\label{fig:medratio}
\end{figure}

In this article we want to concentrate on two medium effect which have been
discussed in detail in ref \cite{SLK95,SKL96,SK96}, namely the reduction of the
pion form factor due to the onset of chiral restoration and the effect of the in
medium pion dispersion relation. As demonstrated in ref. \cite{SKL96} both
effects combine to flatten the invariant mass distribution of dileptons in
qualitative agreement with the CERES data. In ref. \cite{SKL96} the
contributions of other production channels have not been 
taken into account. Also
the effect of the experimental cuts have been ignored thus not allowing for a
quantitative comparison with the data. 
Here we will attempt a first quantitative comparison with the data by including
the above medium effects into our transport model. This is done in the
following way: For each $\pi \pi$ collision we determine the local temperature
from the local covariant pion density 
\be
\rho(x) = \int \frac{d^3 p}{2 \omega(p) (2 \pi)^3} f(x,p)
\ee
where $f(x,p)$ is the pion phase space distribution.  
Assuming local thermal equilibrium we obtain the temperature from
\be
T(r)  = \sqrt{ \frac{\rho(r) (2 \pi)^2}{3}}
\label{eq:loc_temp}
\ee
where $\rho(r)$ denotes the pion density at position $r$. Given the local
temperature we multiply the pion annihilation cross section by the ratio
between in medium and free production as determined in ref. \cite{SKL96,SK96}
(see fig \ref{fig:medratio}).

This procedure involves several approximations: First, we ignore any 
non equilibrium effects on the in medium corrections. This, however, may not be
such a bad approximation, since to leading order in the temperature the in
medium corrections are really effects of the pion density, which we use to
extract the temperature. Secondly, we do not propagate the pions according to
their in medium dispersion relation, but only take its effect into account when
determining the annihilation cross section as described above. This, however,
leads only to a slight difference in the overall reaction dynamics as
demonstrated in ref. \cite{KB93}. Finally, our largest and so far least
controlled approximation is to assume that the pion-form factor does not depend
on its momentum with respect to the heat bath. The determination of the
momentum dependence represents a rather involved computation which we are
currently trying to work out. However, considering the comparatively 
small contribution from the pion-pion annihilation (see below), 
we do not expect that the correction from the true momentum dependent 
form factor will alter our conclusions substantially.

\section{Results and Discussion}
\label{sec:4}
Before we discuss our result in detail let us summarize our findings. By
playing with the initial conditions for the hadronic system we could generate
variations in the dilepton spectra, subject to the CERES acceptance cuts by at
best a factor of two. A large portion of these variations arises from the
contribution of the eta - Dalitz decay, and, therefore, is due to a principally
measurable quantity. 
Considering the the large systematic and statistical
errors of the present CERES data, we unfortunately cannot conclude that these
present data put strong additional constraints on the hadronic equilibration
dynamics. With some initial conditions (not including medium modifications) 
we are able to come within the lower end
of the sum of statistical and systematic error of the CERES data and at the
same time obtain a reasonable agreement with the HELIOS data. We furthermore
find that the in medium modification considered give only an insignificant
enhancement of the total dilepton yield. 
The reason for that is that the pion annihilation, although
important, contributes at best a third of the total yield in the relevant
low mass region. Therefore a factor of two enhancement reduces to less than
30 \% enhancement in the total dilepton spectrum.

\subsection{Dilptons from different inital states}
Let us now turn to the detailed discussion of our results. We have investigated
many different initial conditions and the most representative of those are
listed in table \raf{tab:ini_par}. 
Following the discussion in section \ref{sec:2}
the characterization of the initial hadronic system can be divided into two
major groups
\begin{enumerate}
\item Momentum and configuration space (phase space). This includes 
    choices for the formation time and possible initial radial flow. 
\item Relative abundances of particles
\end{enumerate}

The initial momentum space distribution is controled 
by the choice of rapidty distribution, transverse momentum distribution,
initial flow velocity and, at least for the pions, a possible chemical
potential. The only parameter we used to change the initial configuration space
is the longitudinal extend of the fireball, assume that initially transverse
motion is small. Finally, we can generate possible configuration space and
momentum space correlations by either chosing a finite formation time or,
explicitly, by a finite value of $Y_l$ in eq. \raf{eq:y_z}.

We find that an initial number of about 600 pions (possibly `hidden' in $\rho$
etc mesons) is needed. The relative abundances are either taken as those found
in nucleon-nucleon experiments (indicated by `{\em pp}' in table 
\raf{tab:ini_par})
or as given by chemical equilibrium (indicated by `{\em chem}' 
in table \raf{tab:ini_par}). Whenever we have
chosen chemical equilibrium abundances we also have 
assumed a vanishing formation 
time together with z-y correlations ($Y_l \neq 0$) 
and a slightly more extended fireball (typically $Z_l = 2 \, \rm fm$, except
sets 5 and 6)

\begin{table}[tbh]
\centerline{
\begin{tabular}{c||c|c|c|c|c|c||c|c}
             & Set 1  & Set 2 & Set 3 & Set 4 & Set 5 & Set 6 & Set 7 & Set 8\\ 
\hline \hline
ratios       & p-p    & p-p   & p-p   & therm & therm & therm & p-p   & therm\\
\hline
$N_\pi$      & 350    &  350  &  350  & 114   &  156  &  270  & 1200  & 540  \\
\hline
$N_\rho$     & 70     &  70   &   70  &  96   &  89   &   95  &  240  & 307  \\
\hline
$N_\omega$      & 24     &  24   &  24   &  60   &  62   &   31&  80   & 214  \\
\hline
$N_{a_1}$       &  12    &  12   &  12   &  36   &  25   &   15 & 40   & 86   \\
\hline \hline
$\tau$ [fm/c]   &  0     &   1   &  0.5  &  0    &  0    &   0  & 0.5 &  0   \\
\hline
$T_{init}$ [MeV] &  240   &   200 &  200  &  260  &  190  &  160  & 160 & 190 \\
\hline
$\mu_\pi$ [MeV]   &  0     &   0   &  0    & 130   & 130   &  0  & 0   & 130 \\
\hline
$Y_l$           & 1.6    &  0    &   0   & 1.6   & 1.6   &  1.5  & 0   & 1.6 \\
\hline
$\sigma$        & 0.55   &  1.5  &  1.5  & 0.55  &  0.6  &  0.7  & 1.8 & 0.6\\
\hline
$Z_l$ [fm]      &  2     &  1.5  &  1.5  & 2     &  2.9  &  22   & 2.0 & 3.4\\
\hline
\hline
$Y_l(nucl)$     &   1.7  &   0   &   0   &  1.7  &  1.8  &  1.8  & 0  & 1.8 \\
\hline
$\sigma (nucl)$ & 0.6    &  1.6  &  1.6  &  0.6  &  0.7  &  0.7  & 1.9 & 0.7 \\
\hline \hline
$v_{flow}$      &  0     &  0    &   0   &   0   &  0.35 &  0.4  & 0  & 0.4 \\
\hline \hline
$T_{final}$(prot.) [MeV] &  235  &  230  &  250  & 240   &260  & 250 & 250
  & 290\\
\end{tabular}
}

\caption{Parameter sets used for dilepton calculations for $S+Au$. 
The line labeled `ratios'
indicates whether the relative abundances are taken according to 
proton proton data (p-p) or according to thermal equilibrium (therm). The last
row gives the slope parameter of the final proton spectrum.}
\label{tab:ini_par}
\end{table}

\begin{figure}[tbh]
\setlength{\epsfxsize}{\textwidth}
\centerline{\epsffile{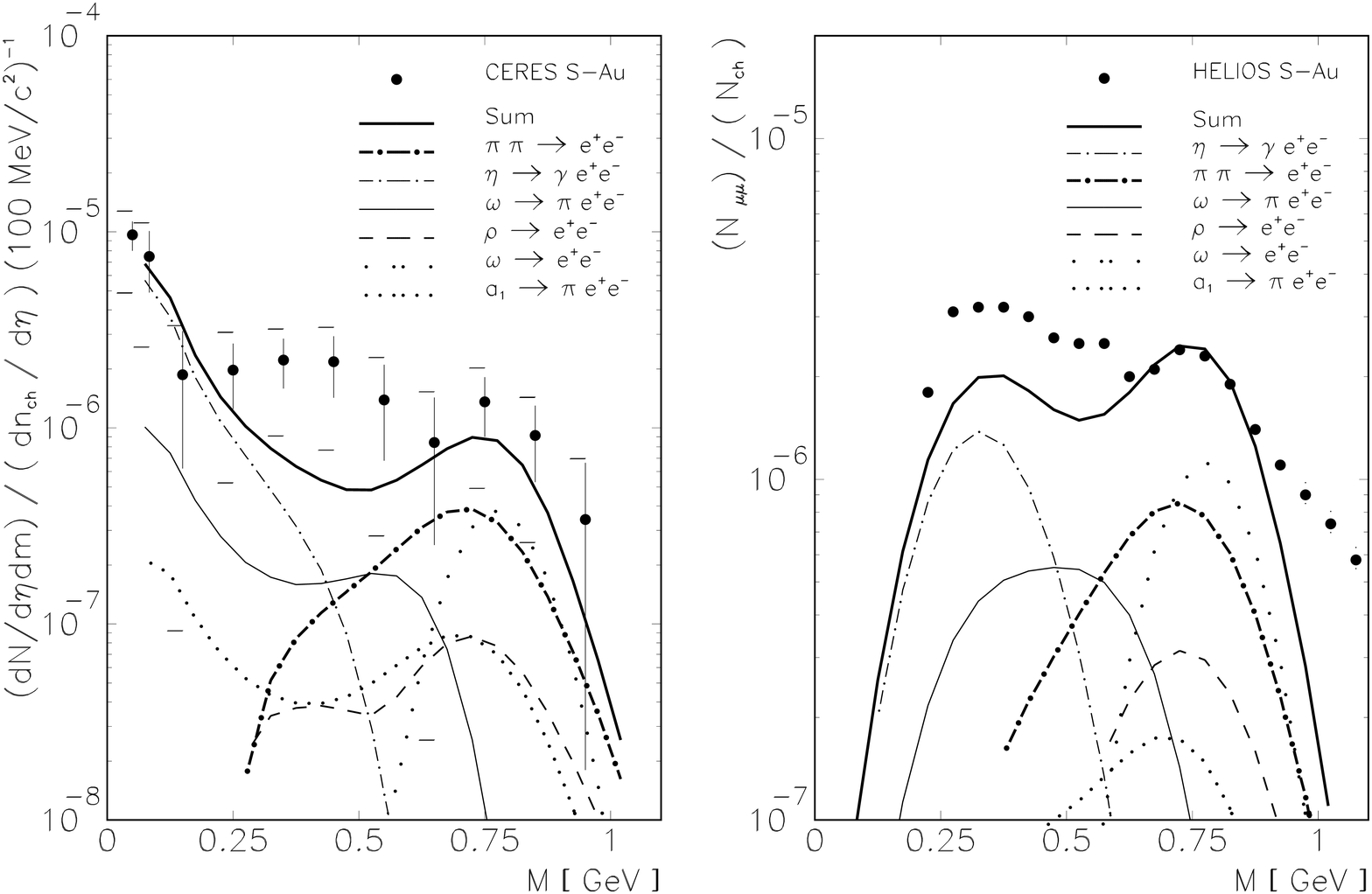}}
\caption{Dilepton production for parameter set 1}
\label{fig:ceres1}
\end{figure}

\begin{figure}[tbh]
\setlength{\epsfxsize}{\textwidth}
\centerline{\epsffile{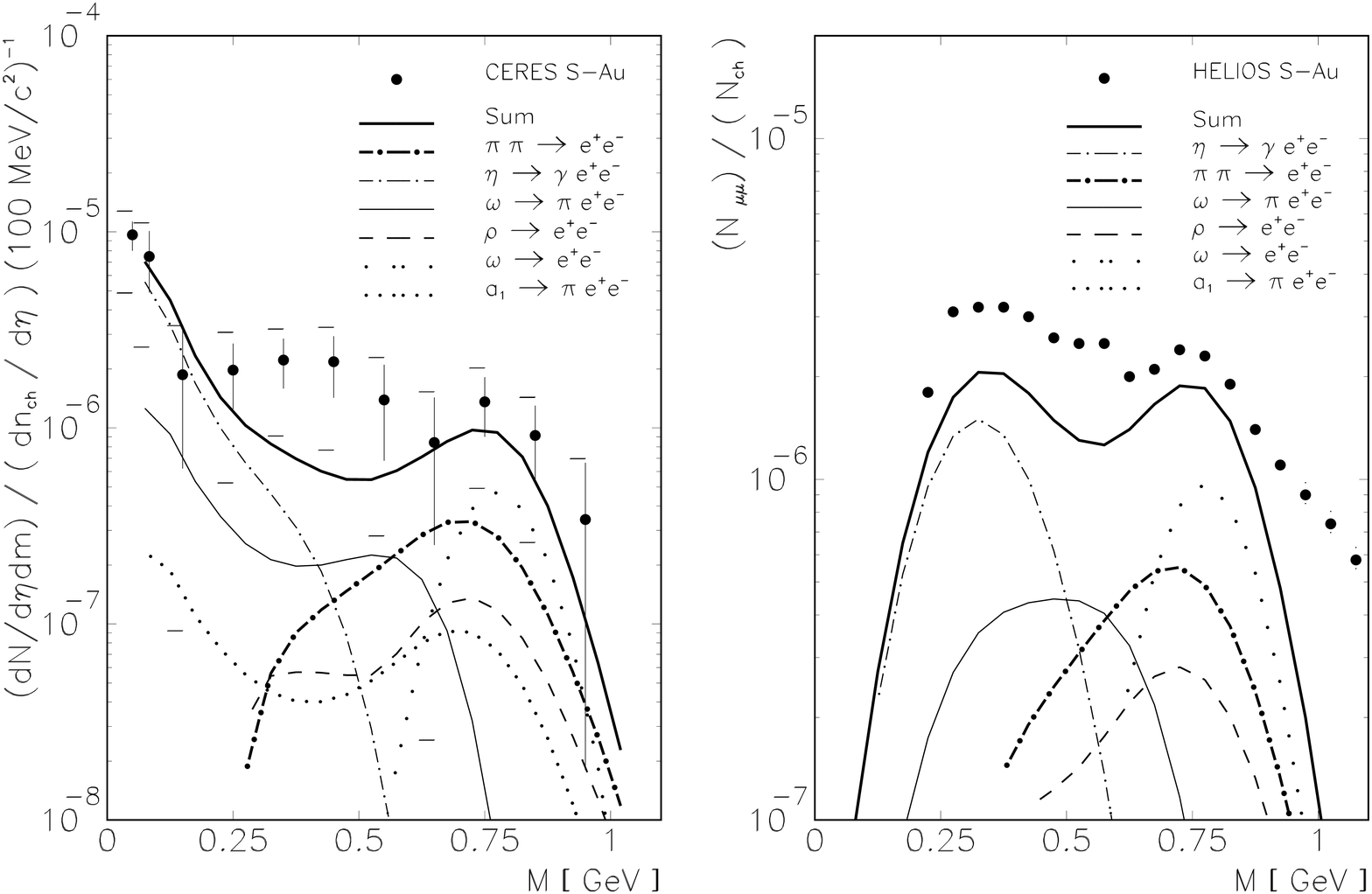}}
\caption{Dilepton production for parameter set 2}
\label{fig:ceres2}
\end{figure}

\begin{figure}[tbh]
\setlength{\epsfxsize}{\textwidth}
\centerline{\epsffile{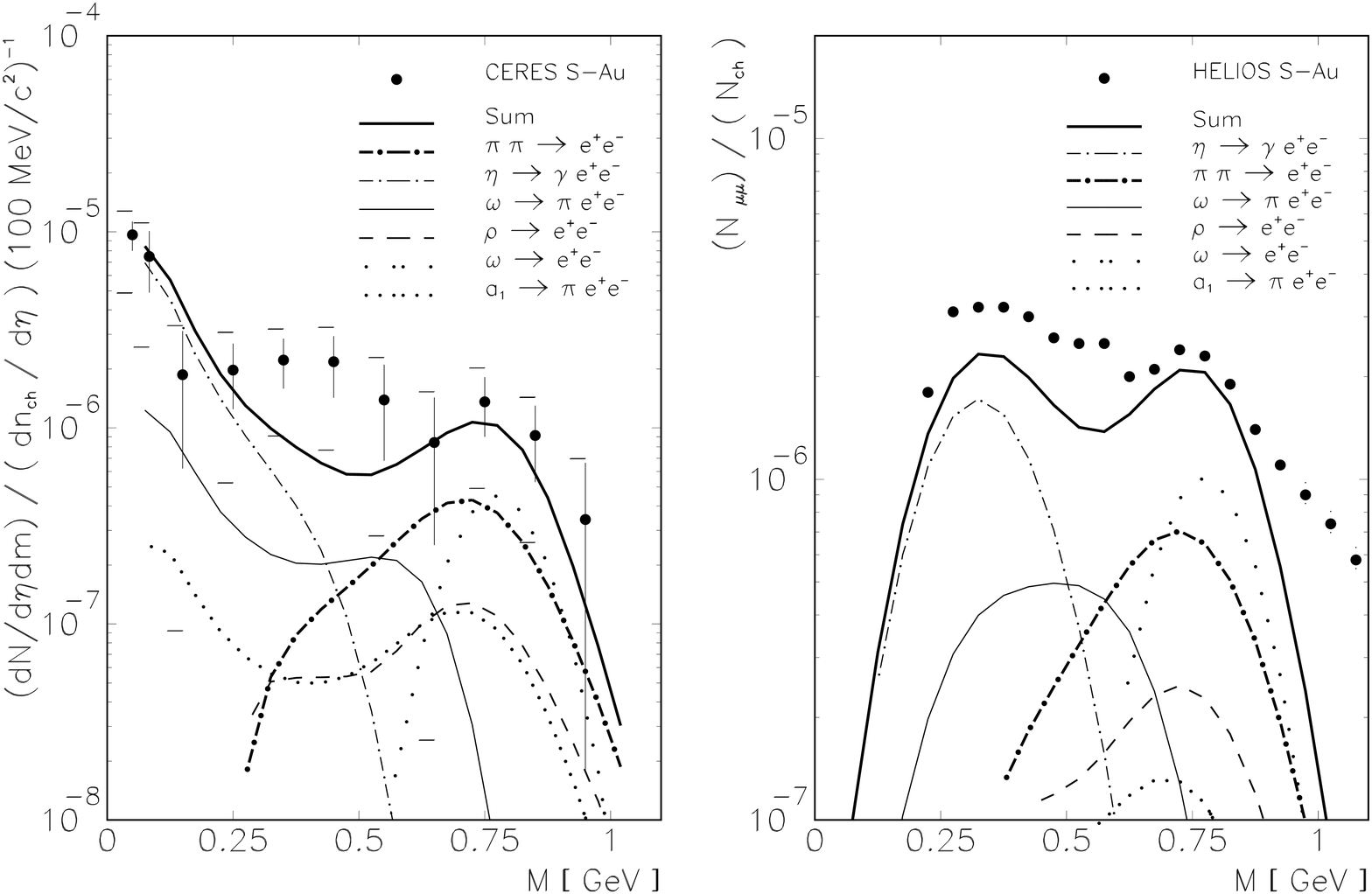}}
\caption{Dilepton production for parameter set 3}
\label{fig:ceres3}
\end{figure}

\begin{figure}[tbh]
\setlength{\epsfxsize}{\textwidth}
\centerline{\epsffile{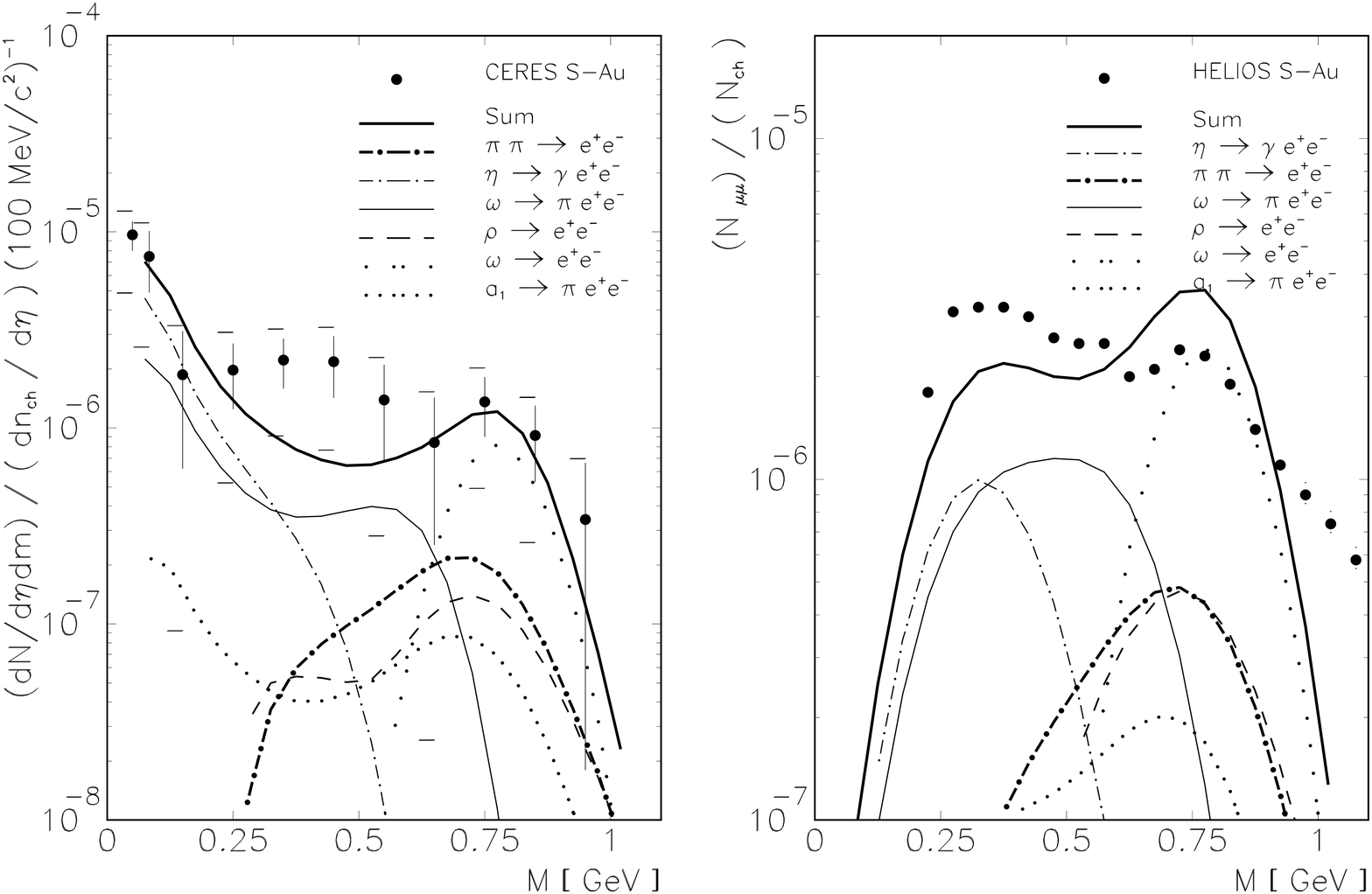}}
\caption{Dilepton production for parameter set 4}
\label{fig:ceres4}
\end{figure}

\begin{figure}[tbh]
\setlength{\epsfxsize}{\textwidth}
\centerline{\epsffile{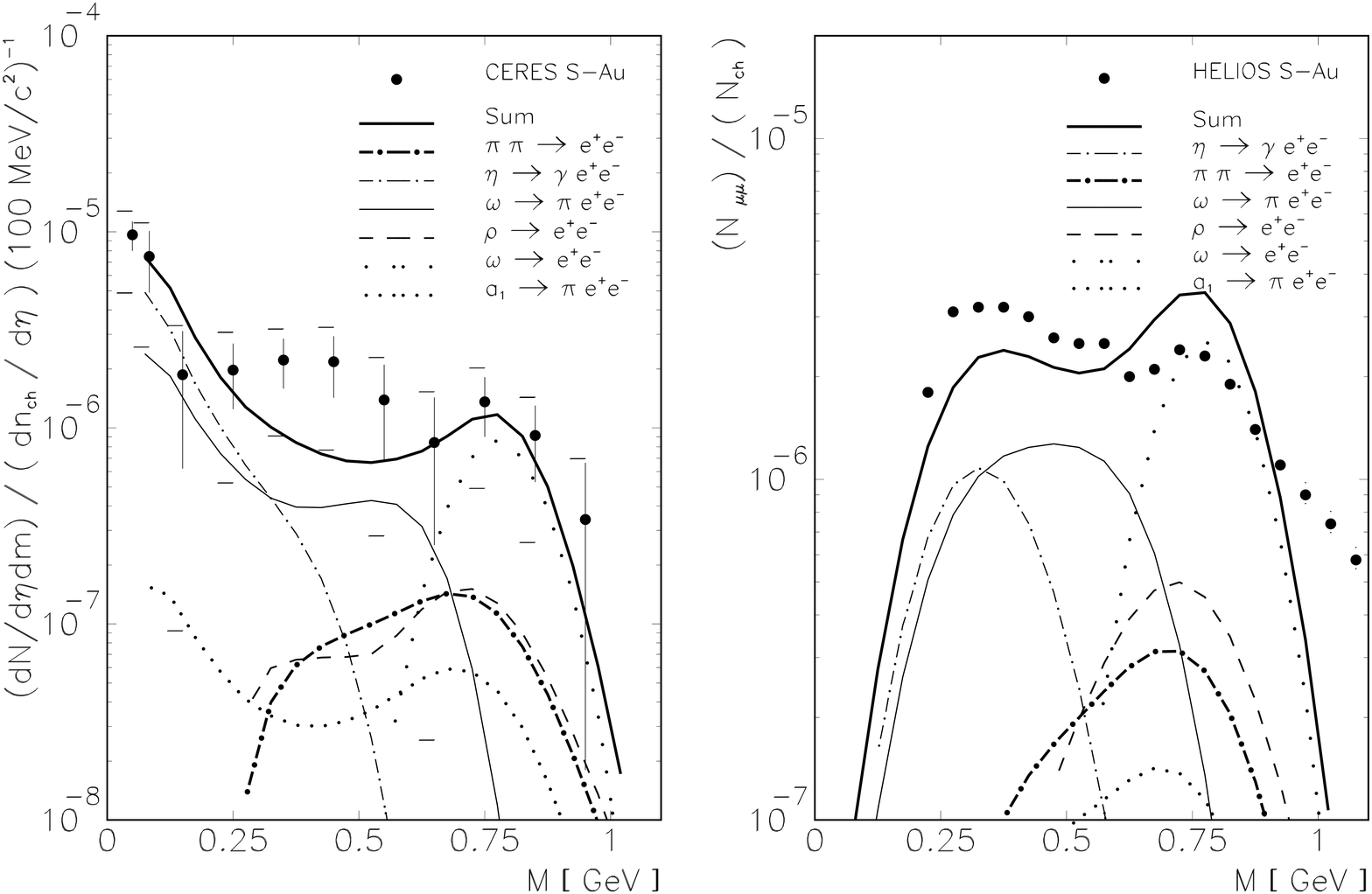}}
\caption{Dilepton production for parameter set 5}
\label{fig:ceres5}
\end{figure}

\begin{figure}[tbh]
\setlength{\epsfxsize}{\textwidth}
\centerline{\epsffile{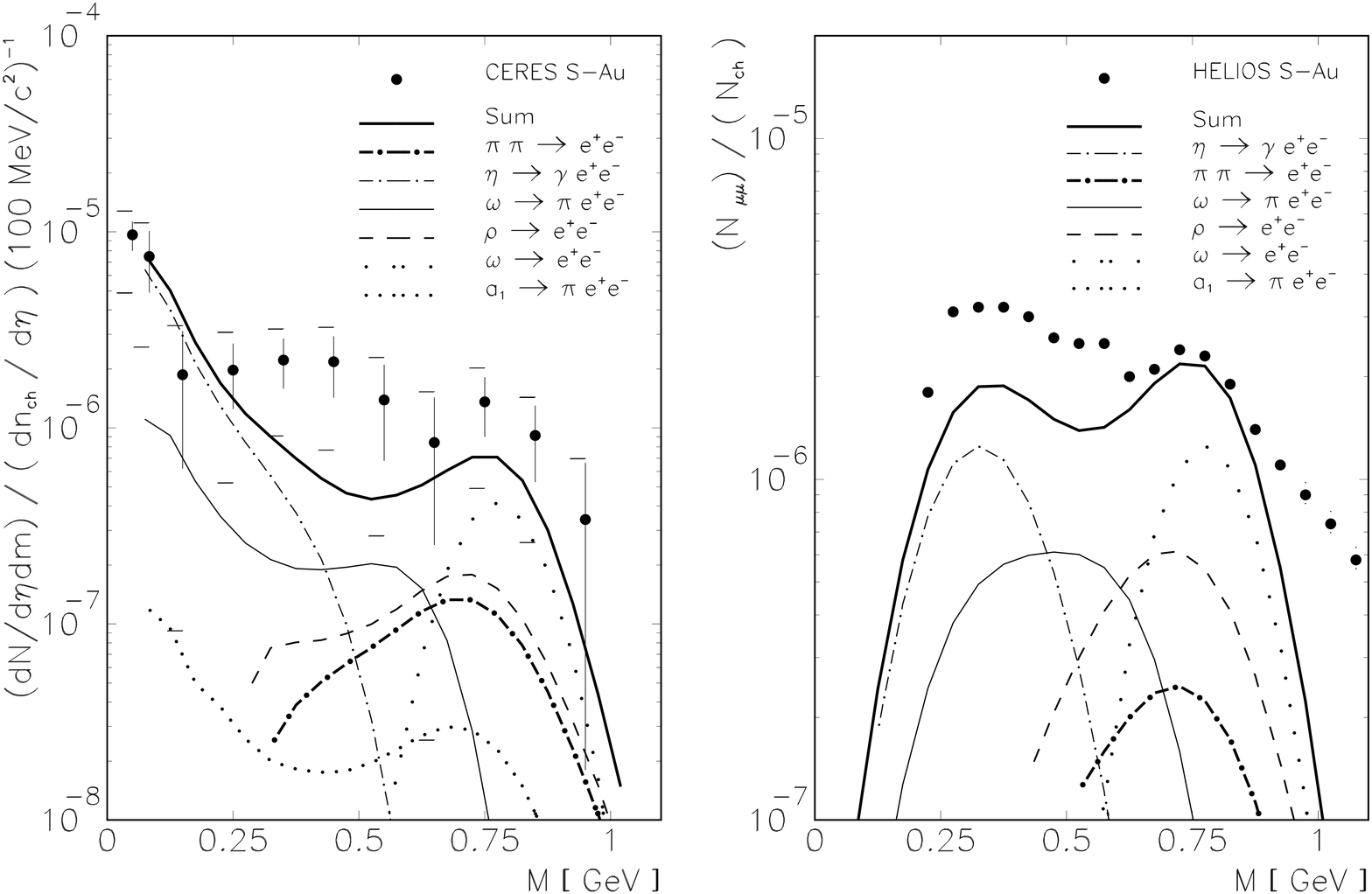}}
\caption{Dilepton production for parameter set 6}
\label{fig:ceres6}
\end{figure}

In figures (\ref{fig:ceres1} -- \ref{fig:ceres6}) we show the resulting 
dilepton invariant mass spectra using the parameter sets given in table 
\raf{tab:ini_par}. To demonstrate the typical agreement we tried to achieve 
with the pion spectra and the proton and pion rapidity distribution the initial
and final distributions for parameter sets 1 and 2 are shown in
fig. \raf{fig:pt_rap1}. For the $p_t$ proton spectra  the slope parameter of
the final spectrum ($T_{final}(prot)$ 
is given in table \raf{tab:ini_par}. These should be
compared with an experimental slope parameter of about $240 MeV$ \cite{NA44}.
Since the protons only indirectly affect the dilepton
production we did not put too much an effort in fine tuning the resulting
proton spectra.

\begin{figure}[tbh]
\setlength{\epsfxsize}{\textwidth}
\centerline{\epsffile{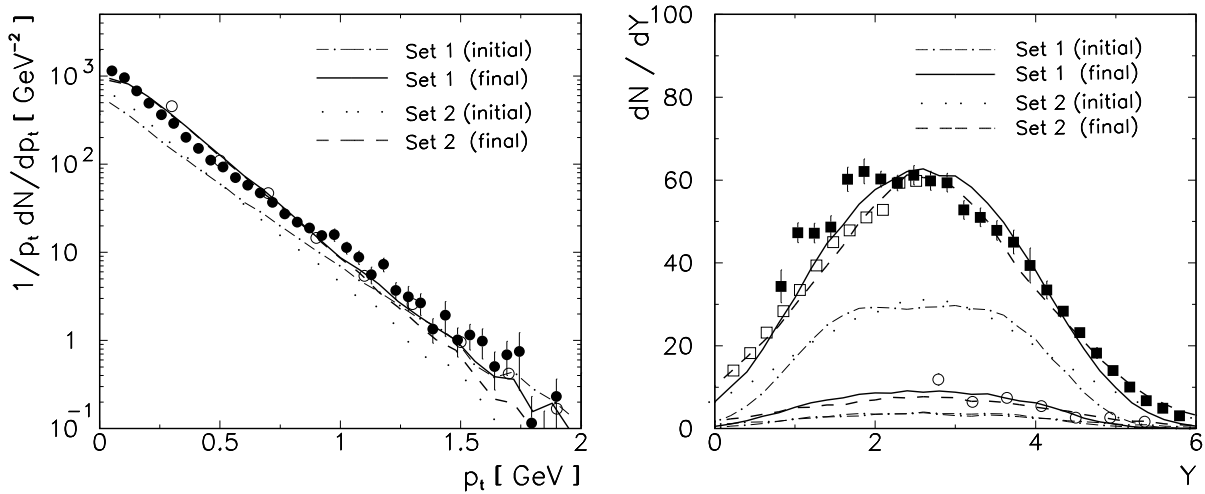}}
\caption{Pion transverse momentum spectrum (left) and pion and proton rapidity
distribution (right) for parameter sets 1 and 2. The experimental data for the
transverse momentum spectrum  are from the NA35 collaboration 
\protect\cite{NA35} (full circles) and the WA80 collaboration 
\protect\cite{WA80_94} (open circles, errorbars left off). The
data for the proton and pion rapidity distributions are from the NA35
collaboration \protect\cite{NA35_2}. The open squares are a reflection of the
datapoints with $Y \geq 2.6$ with respect to $Y = 2.6$.}
\label{fig:pt_rap1}
\end{figure}

Parameter set 1 assumes that at the moment of hadronization the longitudinal
position and the rapidity are correlated ($Y_l \ne 0$, see eq. \raf{eq:y_z} ).
This is
similar to the initial conditions used by Li et al. \cite{LKB95}. The relative 
particle abundances are assumed to be the same as in proton proton collisions.
This leads to about initial 350 pions. With the longitudinal extent of $R_l = 2
Z_l = 4 \, \rm fm$ and a radius of $ 3.5 \, \rm fm$ the resulting initial pion
density is then as large as $\rho_{init} = 2.3 \, \rm fm^{-3}$, which is
certainly larger then one would consider reasonable for a transport
description\footnote{Notice, that due to the initial $z-y$ correlations the
effective pion density is, however, somewhat lower}. 
The initial density is reduced if one uses thermal particle ratios,
which, for the same longitudinal extend, corresponds to parameter set 4. There
we have assumed a fairly large chemical potential of 
$\mu_\pi = 130 \, \rm MeV$. In this
case the pion density reduces to $\rho_{init} = 0.75 \, \rm fm^{-3}$, which is
an acceptable value for a transport description. Notice that the thermal pion
density with that chemical potential and the temperature of $T_{init} = 260 \,
\rm MeV$ would be about twice as high. In this sense parameter set 4 also does
not correspond to initial chemical and thermal equilibrium. 

Initial densities consistent with the thermal ones are considered in parameter
sets 5 and 6. Here we also have allowed for an initial transverse 
flow resulting
in a lower initial temperature. The longitudinal extension of the fireball has
then been 
fixed by requiring that the initial hadron densities agree with thermal
values for the respective chemical potentials. In these cases the initial pion
densities are lower, due to the smaller initial temperature. They are
$\rho_{init} = 0.68 \, \rm fm^{-3}$ and $\rho_{init} = 0.16 \, \rm fm^{-3}$ for
set 5 and 6, respectively

A complementary, and maybe more realistic picture of the initial hadronic
state, is to assume that the creation of a hadron from string fragmentation 
requires a certain time, the so called formation time (see section
\ref{sec:2}). The introduction of a formation time considerably reduces the
initial effective hadron density because only hadrons which have been formed
are allowed to interact. These are essentially only the particles in  the 
same rapidity interval. Parameter sets 2 and 3 have a formation time of 
$\tau = 1 \, \rm fm/c$ and $\tau = 0.5 \, \rm fm/c$, respectively. In this case
we did not give the particles an initial $z-y$ correlation, since our physical
picture is that particles are produced 
independently over the entire volume by the first hard nucleon-nucleon 
collisions. However, due to the finite formation time, for particles which are
allowed to interact  longitudinal position and  rapidity are correlated.
For the longitudinal extent we chose a size of $Z_l = 1.5 \, \rm
fm$, which corresponds to the Lorentz contracted longitudinal size of the
combined sulfur and gold nuclei in the rest frame of the fireball. 
The initial temperature needed to reproduce
the pion spectra is then $T_{init} = 200  \, \rm MeV$. 

Finally we should point out that in all parameter sets except sets 5 and 6 
the relative width of the rapidity distributions (see eq. \raf{eq:y_z} )
between $\pi$, $\rho$ and $\omega$ has been taken from proton proton 
data \cite{ThUl}. These ratios are $\sigma_{\rho}/\sigma_{\pi} = 
\sigma_{\omega}/\sigma_{\pi} = 0.66$. For the $a_1$  we assume a slightly
smaller ratio, namely $\sigma_{a_1}/\sigma_{\pi} = 0.5$.

Parameter sets 1 and 2 (see figures \ref{fig:ceres1} and \ref{fig:ceres2})
give essentially the same results for the CERES dilepton
spectrum. 
The
differences seen in the comparison with the HELIOS data  around
the $\omega$ mass demonstrates the effect of the formation time. In set 2,
where we have a finite formation time of $\tau = 1 \, \rm fm$ the contribution
of the pion annihilation is reduced for the  forward rapidities where the
HELIOS acceptance is located.

The comparison with the calculations seems to suggest that the double hump 
structure in the HELIOS data is mainly due to $\eta$ Dalitz decays at low
masses\footnote{Notice, 
that HELIOS
is measuring dimuons so that the minimum invariant mass is $200 \, \rm MeV$
and, therefore, the rise in the $\eta$ Dalitz 
spectrum does not appear as in the 
CERES spectrum.} and the $\rho$ / $\omega$ decays as well as pion
annihilation at high masses.  
Consequently, possible small discrepancies at low invariant masses
(at the `$\eta$-hump') should only be taken seriously once a precise
measurement of the $\eta$ abundance at forward rapidities is available. Without
this information the disagreements seen in our results can always be `fixed' by
fine tuning the forward pion rapidity distribution and the hard end of the pion
transverse momentum spectrum, since we use $m_t$ scaling to obtain the $\eta$
mesons from the final pion distribution. 

Reducing the formation time to $\tau = 0.5 \, \rm fm/c$ (Set 3, figure
\ref{fig:ceres3} ) increases
the contribution from the pion annihilation, bringing our results close to the
lower end of the sum of statistical and systematic error bars of the CERES data
in the interesting mass region around $400 -500 \, \rm MeV$. 
The agreement with the HELIOS data is pretty reasonable in this case except
around an invariant mass of about $500 \, \rm MeV$, where we are about a factor
of two too low, even if we fine tuned the $\eta$ Dalitz contribution. 
In this case also the
contribution of the $\eta$-Dalitz is slightly larger then in Set 1. Because of
the shorter formation time more flow is generated which in turn gives a
slightly harder pion spectrum (which, however, is still in very good agreement
with the data) and thus a few (20 \%) more $\eta$ mesons.

The one single process which contributes directly to the interesting region
around $M \simeq 400-500 \, \rm MeV$ is the Dalitz decay of the $\omega$ meson.
Since the creation cross section of $\omega$ meson via $\pi$-$\rho$ fusion is
rather small the only way to increase the number of $\omega$ mesons is by
assuming thermal particle ratios with a large pion chemical potential. This is
done in parameter set 4, where we have assumed a pion 
chemical potential $\mu_\pi = 130 \, \rm MeV$. Comparing with parameter set 1,
this clearly increases the dilepton yield in the relevant mass region and the
main contribution is from the $\omega$-Dalitz decays. At the same time 
also the contribution of the direct decay of the $\omega$ is
increased, leading to an overprediction of the HELIOS data in the
$\rho$-$\omega$ mass region.

As already mentioned, for parameter sets 5 and 6 we have adjusted the
longitudinal size of the fireball such that the particle density agrees with
the thermal one. We have also allowed for initial radial flow and, as a result,
the initial temperature is smaller. Furthermore, unlike in the previous cases,
the widths of the rapidity distributions have been taken to be the same for all
mesons. Assuming thermal densities, although
apparently more consistent, reduces the hadron dynamics to some extent, in
particular in case of set 6, where we have a longitudinal size of $44 \, \rm
fm$. These are essentially freeze out conditions 
and it appears rather unlikely that
hadrons do not interact and/or have expanded in the radial
direction before that. 
In case of set 5 the longitudinal extent is not much larger than in
the comparable set 1 and this may be a more realistic scenario, where hadrons
are created in thermal equilibrium. Again, because of the large pion chemical
potential in set 5, there is a large fraction of initial omegas. Consequently,
we find a sizeable contribution in the interesting mass region and an
overshooting in the $\rho$ $\omega$ region in case of the HELIOS data.

\begin{figure}[tbh]
\setlength{\epsfxsize}{\textwidth}
\centerline{\epsffile{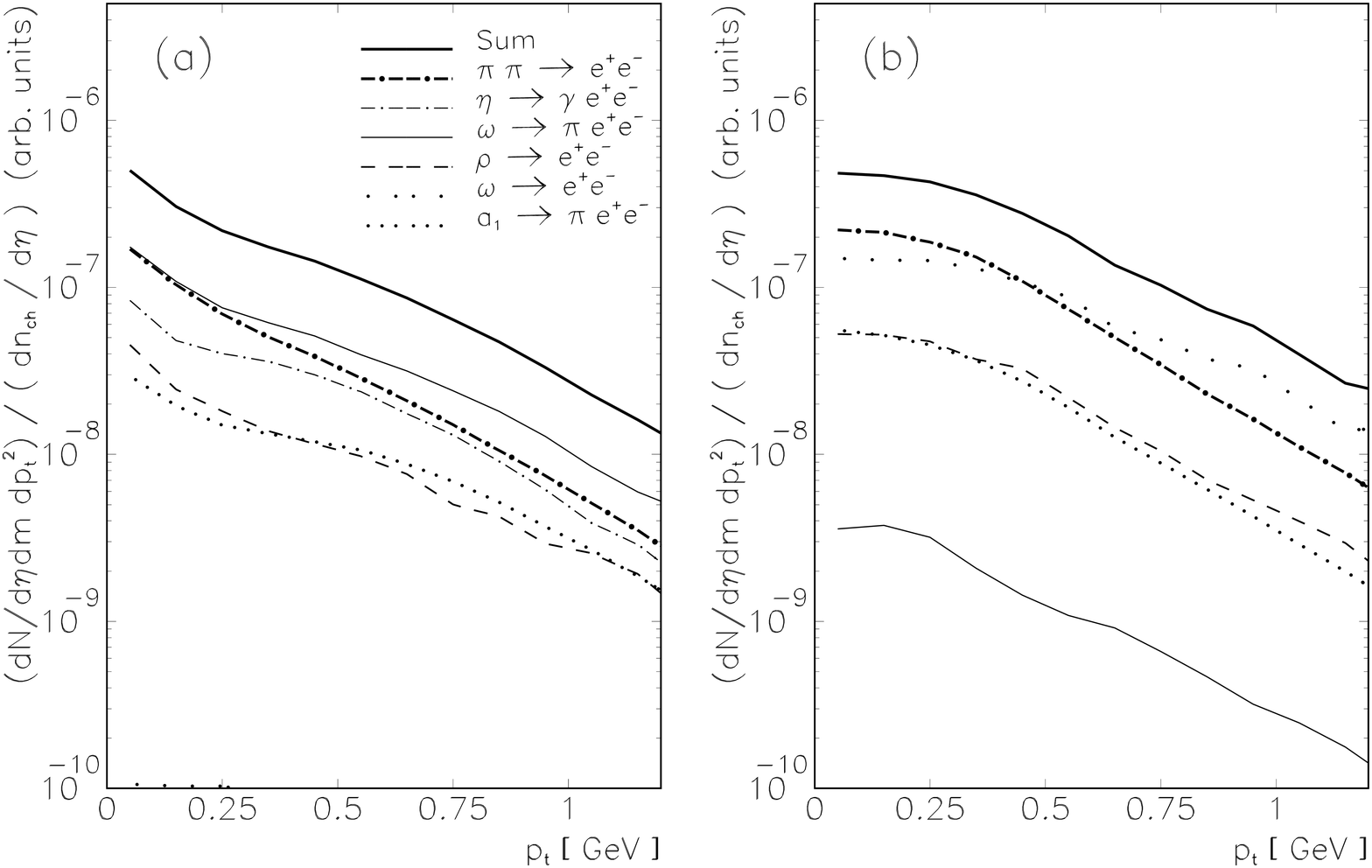}}
\caption{Transverse momentum spectra for dileptons using the CERES acceptance
and parameter set 3. The ranges in invariant mass are 
$400 \, \rm MeV \leq M \leq 450 
\, \rm MeV$  (a) and $750 \, \rm MeV \leq M \leq 800 \, \rm MeV$ (b)}
\label{fig:ceres3_pt}
\end{figure}

Additional information may be obtained by studying transverse momentum spectra
of the dileptons once data of sufficient statistics are available. In figure
\ref{fig:ceres3_pt} we show the resulting transverse momentum spectra using the
parameter set 3 for the invariant mass regions $400 \, \rm MeV \leq M \leq 450
\, \rm MeV$  (a) and $750 \, \rm MeV \leq M \leq 800 \, \rm MeV$ (b), 
respectively. We do not observe any significant structure in the spectrum and
all contributions look more or less exponential. Only in the $\rho$ - $\omega$
mass range (b) the contribution of the omega decay shows a significantly larger
slope parameter (`temperature'). This is essentially a result of the transverse
flow which is generated during the expansion. This affects mostly the
$\omega$ mesons,  which decay predominantly after freeze out, i.e. after the
transverse flow has been build up. Pion annihilation, the other strong
contribution in this mass range, on the other hand, contributes mostly at the
initial stage of the the expansion, and thus is not affected by flow.
Therefore, already with the present mass resolution of the
the CERES detector, the relative importance of the omega decays 
can be determined by  a careful measurement of the slope parameter 
of the dilepton transverse momentum spectrum. This will then provide a crucial
constraint to available model calculations.

To conclude this part, it appears that without any in medium effects we are
able to more or less reproduce, within errors, the CERES data as well as the 
HELIOS data. Except for the first three data points, our
`best' result, set 3, seems to be consistent with the CERES data, if the points
are shifted towards the lower end of the systematic error. Looking at the
HELIOS and CERES data together, there seems to be some enhancement at an
invariant mass slightly above $500 \, \rm MeV$. However, given the large
systematic errors of the CERES data, and the fact that the systematic errors
of the HELIOS data are not given, it seems to be too early to declare the
presence of any in medium effects. If, of course, an improved measurement
would confirm the central values of the CERES data, from our
investigations, we do not expect that any non-equilibrium effects of the
hadronic phase would give that many dileptons in the relevant invariant mass
region around $400 - 500 \, \rm MeV$.
 
We have demonstrated that different initial hadronic configurations can lead to
differences in the dilepton yield while still consistent with the hadronic
observables. It also seems that the HELIOS data rule out initial 
partical ratios derived from chemical equilibrium based on a large pion 
chemical potential. Unlike the $\rho$, the decay of the $\omega$ into dileptons
is not expected to be reduced as a result of the resoration of chiral symmetry
\cite{LSY95}. Also, a possible collisional broadening 
 of the omega would not change this,
because the experimental resolution is already much larger than any expected 
in medium width of the omega \cite{Hag95}.

From all the many configuration we have investigated (also
those not reported here) it appears, however, 
that it is very unlikely to go above the
lower end of the combined error of the CERES data the  mass region around
$400 - 500 \, \rm MeV$. 
As we shall discuss in the following
section, we also do not expect that  our proposed in medium correction would
improve the situation considerably.

\subsection{Medium effects}
\begin{figure}[tbh]
\setlength{\epsfxsize}{\textwidth}
\centerline{\epsffile{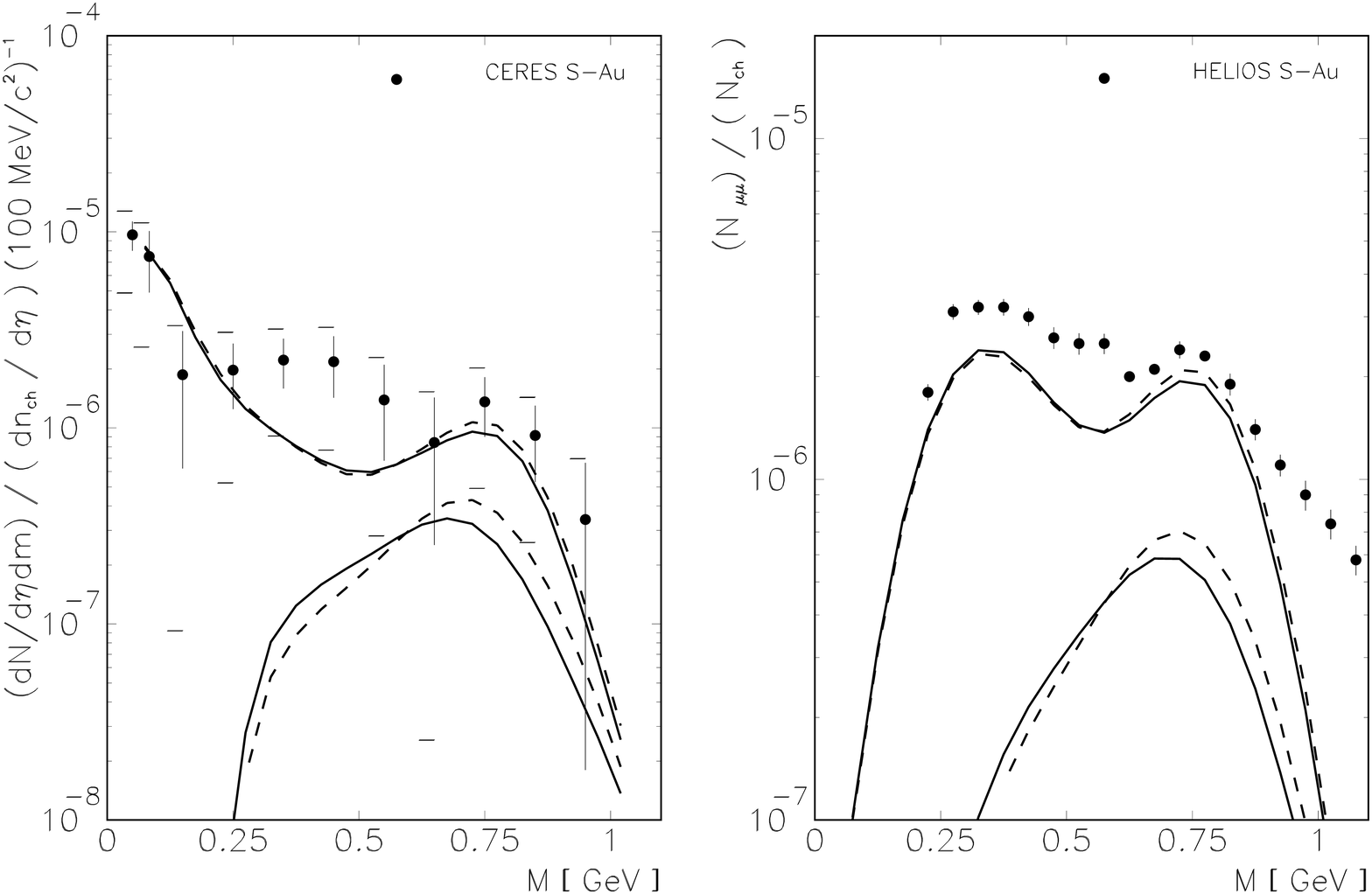}}
\caption{Dilepton production for parameter set 3 with (full lines) and 
without (dashed lines) in medium 
modification of the pion annihilation cross section.}
\label{fig:ceres3_med}
\end{figure}

Let us now turn to the results including the proposed in medium modification of
the pion annihilation cross section (see section \ref{sec:3b}). Since our in
medium modification only affects the pion annihilation let us consider the
parameter set 3, which has the biggest contribution from this channel. This then
gives an upper limit of what we should expect for the different scenarios
considered here. The resulting dilepton spectra are shown in figure
\ref{fig:ceres3_med} together with the results without in medium effects, c.f.
figure \ref{fig:ceres3}. We only show the total yield and the contribution from
the pion annihilation. All other contributions  are identical to those in
figure \ref{fig:ceres3}, since the way we implement the in medium corrections
does not affect the expansion dynamics. While there is an  enhancement in
the pion annihilation contribution at low invariant masses and a suppression
around the $\rho$-$\omega$ mass, the effect in the total spectrum is hardly
visible, especially in the interesting mass region. This is simply due to the
fact that the pion annihilation contributes less than 1/3 to the total yield in
this region and even an enhancement of a factor of two would increase the total
spectrum by less than 30 \%. 

Comparing with the changes in the dilepton yield we could generate by altering
the initial conditions for the hadronic phase, the effect of the in medium
corrections is small. We, therefore, have to conclude that it is 
probably impossible to learn something about the in medium pion dispersion
relation from SPS-energy dilepton experiments unless the 
dynamics of the hadronic phase is
extremely well known, and of course, much more precise measurements are
feasible.

\subsection{Prediction for Pb+Pb}

\begin{figure}[tbh]
\setlength{\epsfysize}{3in}
\centerline{\epsffile{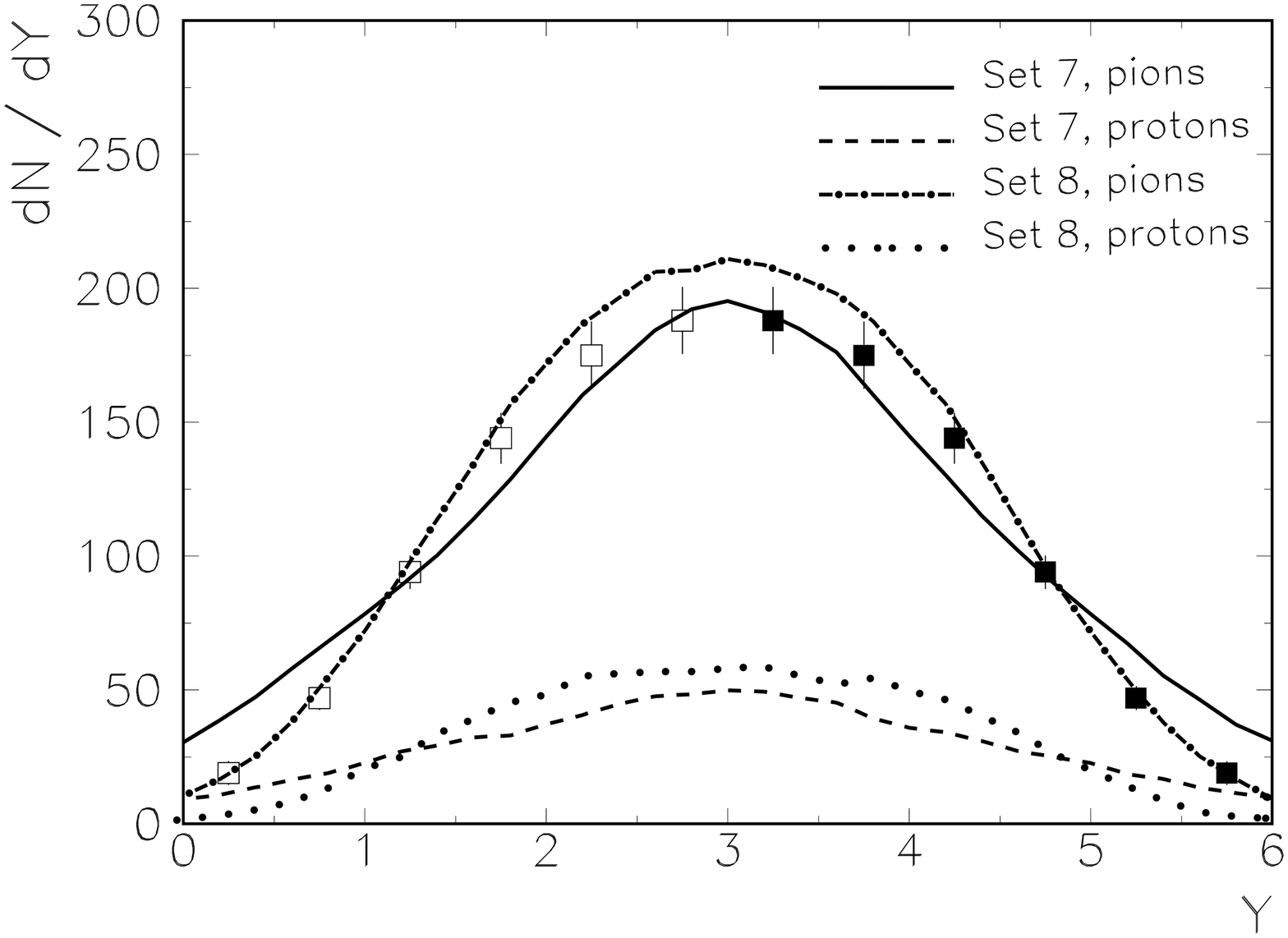}}
\caption{Pion and proton rapidity distribution for Pb+Pb. Data points are
preliminary results for negatively charged hadrons from the 
NA49 collaboration.}
\label{fig:rap_pb}
\end{figure}

Currently, the system $Pb+Pb$ at $150 \, \rm GeV/A$ is investigated by the
CERES collaboration so that it seems useful to try to make some predictions on
what we would expect for the dilepton yield of this large system. Within our
approach, this can only be a rough estimate, since final hadronic data for this
system are not yet available. However, preliminary results for the rapidity
distribution of negatively charged particles from the NA49 collaboration
\cite{NA49} as well as preliminary spectra from the NA44 collaboration
\cite{NA44} are
available. NA44 reports that the slope parameter for pion transverse momentum
spectrum  changes only  little when going from S+Pb to Pb+Pb. We, therefore,
require that the final pion spectrum agrees with the S+Au data from the NA35 
collaboration up to an overall constant. Our resulting rapidity distributions
are shown in figure \ref{fig:rap_pb} together with preliminary data from the
NA49 collaboration \cite{NA49}.

\begin{figure}[tbh]
\setlength{\epsfxsize}{\textwidth}
\centerline{\epsffile{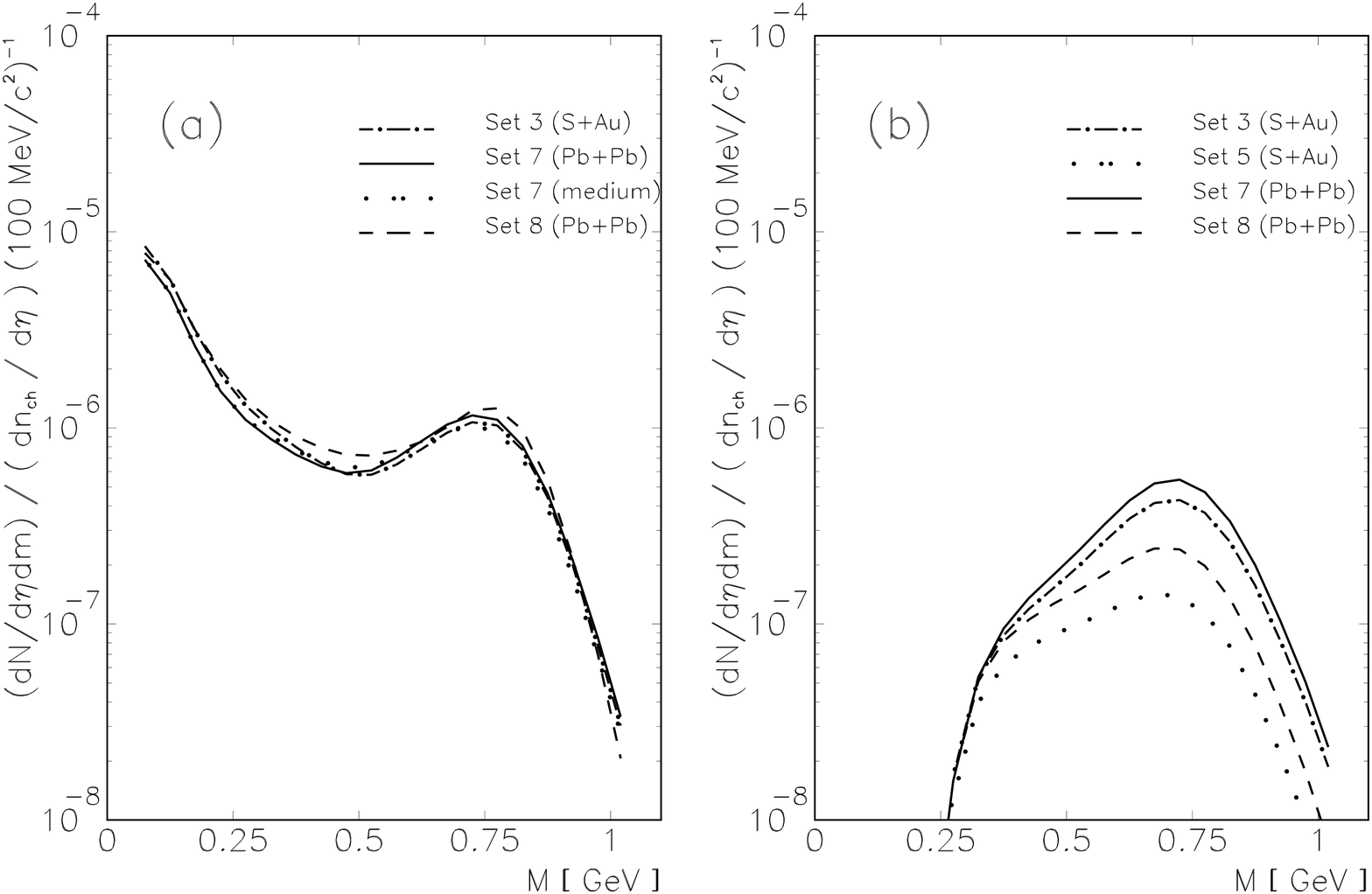}}
\caption{Prediction for Pb+Pb. (a) Total yield, (b) contribution from pion
annihilation} 
\label{fig:cerespb1}
\end{figure}

In figure \ref{fig:cerespb1}(a) 
we show our prediction for the Pb+Pb result, based
on the CERES acceptance and normalization, i.e. dileptons per charged
particles. The parameterizations we have used are given in table 
\ref{tab:ini_par} (sets 7 and 8). They follow the same philosophy as sets 3 and
5 for the S+Au case, respectively. 
In order to model  a hadronic system generated in Pb+Pb collisions we
increased the transverse radius to $6.5 \, \rm fm$ and the number of total
pions ($\sim 2000$) and baryons (400). 
The relative abundances are the same as in sets 3 and 5, respectively.

For a comparison, we also show the results
for S+Au obtained with parameter set 3. We find, that if normalized to the
number of charged particles, the dilepton spectrum hardly changes by going from
S+Au to the heavier system of Pb+Pb. This result may be surprising at first
sight, because in the heavier system many more pions are created which should
give rise to more dileptons from pion annihilation. However, one has to be a
little careful with this argument. The rate of dilepton production from pion
annihilation is roughly proportional to the square of the pion {\em density}
\be
\frac{d N_{e^+e^-}}{V \tau} \sim \rho_\pi^2
\ee
where V is the the volume and $\tau$ is the lifetime of the fireball and
$\rho_\pi$ the average pion density. 
Consequently, the total number of dileptons produced divided by the number of
charged particles (mostly pions) is {\em linearly} proportional the average 
pion density
\be
\frac{N_{e^+e^-}}{N_{charged}} \sim \frac{\tau V \rho_\pi^2}{N_{charged}}
\sim \tau \rho_\pi
\ee
Here we assumed that $N_{charged} \sim N_\pi$. It is probably fair to assume
that the volume of the fireball scales like the volume of the initial nuclei
and, therefore, more or less like A, the number of nucleons. But also the
number of produced pions (negatively charged hadrons) seems to scale the same
way. NA35 reports about 100 negative pion produced in S+S collisions
\cite{NA35_2} and NA49 a preliminary number of about 700 in Pb+Pb collisions
\cite{NA49}. Consequently, since both the volume and the number of pions seem
to scale more or less with the size of the system, the average pion density
reached in these collisions is essentially independent of the system size.
Remains the lifetime of the fireball, $\tau$, which is probably larger
for a bigger system but depends on the detailed expansion dynamics. 
The effect of the lifetime is demonstrated in figure \ref{fig:cerespb1}(b),
where the contributions of the pion annihilation is plotted. If we compare the
results for S+Au (set 5) with those for Pb+Pb (set 8), we find about a factor
of 2 increase around the $\rho$-$\omega$ mass. In both calculations the initial
pion density is roughly the same, so that difference comes from the longer
lifetime of the larger system.

Notice, that the CERES experiments accepts
dileptons in the rapidity interval $2.1 \leq Y \leq 2.65$ and, thus, favors the
S+Au system, which has a fireball rapidity of about $Y = 2.6$, whereas 
Pb+Pb is located at $Y = 3$. 
This difference in acceptance is also the reason why the increase in the pion
annihilation contribution is not reflected in the total yield, especially
around the $\rho$-$\omega$ mass. Dileptons from $\omega$ and initial $\rho$
mesons, which scale like the number of charged particles, contribute less.

Again, our proposed in medium modification due to the pion dispersion relation
leads only to a small correction (see dotted curve in 
figure \ref{fig:cerespb1}(a) ).

To summarize our prediction, if there are no in medium effects such as dropping
$\rho$ masses, we expect that the dilepton spectrum for Pb+Pb 
subject to the CERES acceptance and normalization should not differ 
significantly from that for S+Au. If the statistical and systematic errors are
smaller for the Pb+Pb data, as expected \cite{Tserruya}, we would actually
expect that in the mass region of $400-500 \, \rm MeV$ 
the central values should come down somewhat as compared to the published S+Au
data. If, on the other hand, the Pb+Pb data show a significant increase over
S+Au, some additional, prehadronic 
production mechanism or strong in medium modifications 
such as dropping vector meson masses etc. would be called for.

\subsection{Baryon decays}
In our calculation we have not taken into account the possible Dalitz decays of
any baryons. While we have taken into account the Delta resonance for the
expansion dynamics we have not included its Dalitz decay mainly because it will
only contribute to invariant masses lower than $300 \, \rm MeV$. However, at 
the temperatures considered also heavier baryon resonances will be present. 
For a temperature of 180 MeV we find that about 12 \% of the baryon number will
be in states heavier than the Roper resonance ($N^*(1440)$), such as the
$N^*(1520)$, $N^*(1535)$, the $\Delta(1600)$, and the $\Delta(1620)$. 
While the
$N^*(1440)$ hardly couples to photons (the branching ratio is smaller than a
tenth of a percent) the higher nucleon and Delta states have considerable
photon decay width. Also their masses are large enough such that the
Dalitz decays could contribute to the relevant region around $400 \, \rm MeV$.
Since the lifetime or these resonances is rather short ($\Gamma \geq 150 \, \rm
MeV$) the dominant contribution to the dilepton spectrum will come during the
lifetime $\tau$ of the fireball, where we may assume a constant number of these
resonances. The total contribution to the dilepton yield, therefore, is
\be
N_{e+ e-}^{baryons} = \tau N_{baryons} \Gamma_{e^+e^-}^{baryons}  + 
     N_{baryons} \frac{\Gamma_{e^+e^-}^{baryons}}{\Gamma_{tot}^{baryons}}
\ee
The second term takes into account the contribution after freeze. It can
be ignored for our estimate 
since the lifetime of the resonances $1 / \Gamma_{tot}^{baryons}$ 
is much smaller that that of the fireball ($\tau$).
 
To give an upper limit for the contribution we assume a lifetime of 
$\tau = 10 \, \rm fm/c$ for the fireball and that the heavy baryons constitute 
25 \% of the total baryon number (which is a factor of {\em two} more than
chemical equilibrium would suggest at a temperature of $180 \, \rm MeV$).
For the Dalitz decay width $\Gamma_{e^+e^-}^{baryons}$ we use the formula (4.8)
of
Wolf et al. \cite{WBC90}, which has been derived for the Dalitz decay of the
Delta of a given mass $M$. The coupling we adjust by assuming that the partial
width into photon plus nucleon is $1 \, \rm MeV$, which again is on the large
side. For the mass we chose $M = 1.6 \, \rm GeV$ which is in 
between the nucleon and
Delta excited states under consideration. 
Finally, as done in all calculations, we assume that there are 60 baryons in
the fireball, so that we are dealing with 15 high mass baryons.

As a reference we use the Dalitz
decay of the $\omega$ meson, since its contribution can also be simply
estimated. Due to its small decay width, most  $\omega$ mesons 
decay after freeze out and, therefore, their contribution is given by
\be
N_{e+ e-}^{\omega} = N_{\omega} 
\frac{\Gamma_{e^+e^-}^{\omega}}{\Gamma_{tot}^{\omega}}
\ee

In fig. \raf{fig:baryon} we show the ratio between 
Dalitz decays from the baryon and that of the $\omega$ mesons, assuming that
there are 24 $\omega$ mesons at freeze out. This number is comparable with our
calculation presented in fig \raf{fig:ceres1}. 
We see that even with our optimistic assumption the contribution of the baryons
to the dilepton spectrum is less the half of that from the Dalitz decay of the
$\omega$. It is probably fair to assume that the experimental acceptance cuts
apply equally to both. By comparing with the result shown in fig.
\raf{fig:ceres1} we thus conclude that the contribution of the baryon Dalitz
decays only lead to a small correction $\simeq 10  \% $ to the results 
presented above, which may be slightly larger but certainly not significant in
case of Pb+Pb.

\begin{figure}[tbh]
\setlength{\epsfysize}{4in}
\centerline{\epsffile{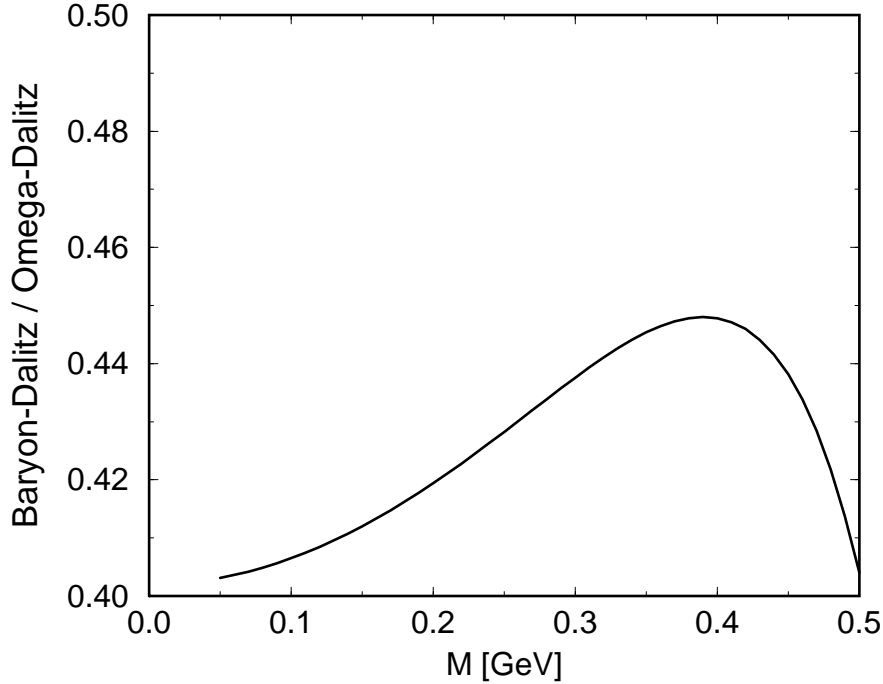}}
\caption{Ratio of baryon-Dalitz decay over $\omega$-Dalitz decay}
\label{fig:baryon}
\end{figure}

\section{Conclusion}
In this article we have studied dilepton production form a hadronic system
believed to be created in an SPS energy heavy ion collisions. We have addressed
the question to which extent dilepton data provide information on the
equilibration dynamics in the hadronic phase. To this end we have investigated
several different initial conditions for the hadronic phase, subject to the
constraint that the final hadronic observable agree with the data. The
comparison with the dilepton data could rule out an initial state where the
relative abundances of hadrons is given by chemical equilibrium based on a
large pion chemical potential. Variations of the initial momentum and
configuration space distributions affected the dilepton  invariant mass
spectrum by not more than a factor of two, which is small compared to the error
bars of the data. Therefore, with the presently available data, the question
of the equilibration dynamics can not be settled. 

We were not able to find a scenario which would give enough dilepton in the
mass region around $400 - 500 \, \rm MeV$ to reach the central values of the
CERES data. Our `best' results agrees with the lower end of the sum of
statistical and systematic errors in the region. Considering the large
experimental uncertainties we can not conclude that any in medium modifications
are needed in order to explain the present data. 

We predict essentially the same dilepton spectrum for the larger system Pb+Pb
as for S+Au, provided the CERES acceptance and normalization are used. 
Since for
this system smaller experimental errors are expected, these data may give very
valuable information about possible in medium corrections such as dropping
vector meson masses. 

We have investigated the effect of an in medium pion dispersion relation
and pion form factor on the dilepton spectrum. While the effect is visible in
the pion annihilation channel, it hardly affects the total dilepton yield.
Therefore, unfortunately it seems rather unlikely that the presence of an in
medium pion dispersion relation can be detected in SPS-energy heavy ion
collisions.

Our conclusions could be much more sharpened if 
precise measurements of the  the
$\eta$ mesons and the $\omega$ mesons would be available. The latter, of
course, could be obtained best with a considerably improved mass resolution 
of the dilepton
spectrometer. 
However, a careful measurement of the transverse momentum spectrum in the
$\rho$ - $\omega$ mass region would already provide some information on the
relative importance of the omega decays. 
Further constraints could be obtained by looking at  pion
correlation (HBT) data. These may provide valuable information about the
space-momentum  correlations and thus restrict our choices of initial 
conditions. However, considering, the rather weak sensitivity of the dilepton
data with respect to the variation of the initial phase space, 
we do not expect much additional insight from such an investigation.

\ \\
\ \\
\noindent
{\bf Acknowledgements:}
We would like to thank the CERES collaboration, in particluar A. Drees,
I. Tserruya and Th. Ullrich for many useful discussions concerning their data.
We would further like to thank the NA35 collaboration and in particular
G. Odyniec for providing the data files for the hadronic spectra.
V.K. would especially like to thank G.Q. Li for aggreeing to and helping in 
a detailed comparison of our transport models. 
Helpful discussions with G.E. Brown and C.M. Ko are also acknowledged.

%\bibliography{/home/vkoch/tex/bibl/pion}

\end{document}